\newcounter{secnum}

\newcommand{\mysection}[1]{%
\vspace{1.25\baselineskip}
\stepcounter{section}
\stepcounter{secnum}
\centerline{\large\bf\thesecnum. #1}
\vspace{1pt}
}

\newcounter{subsecnum}[secnum]

\newbox\slashbox
\newdimen\slashwd
\newcommand{\slashed}[1]{%
\setbox\slashbox=\hbox{$#1$}%
\slashwd=\wd\slashbox%
\hbox to\slashwd{\hss/\hss}%
\llap{$#1$}}

\documentstyle[12pt]{article}

\begin{document}
 
\begin{titlepage}
\begin{center}
{\bf The Standard Model in the Alpha Gauge is Not Renormalizable}\\
\end{center}

\begin{center}
Hung Cheng$^1$\\
Department of Mathematics, Massachusetts Institute of Technology\\
Cambridge, MA  02139, U.S.A.\\

\bigskip
and
\bigskip

S.P. Li$^2$\\
Institute of Physics, Academia Sinica\\
Nankang, Taipei, Taiwan, Republic of China\\
\end{center}



\noindent
1. E-mail: huncheng@math.mit.edu

\noindent
2. E-mail: spli@phys.sinica.edu.tw
 
\end{titlepage}

\begin{abstract}
  We study the Ward-Takahashi identities in the standard model
  with the gauge fixing terms given by (1.1) and
  (1.2).  We find that the isolated singularities of the 
  propagators for the unphysical particles are poles of even
  order, not the simple poles people have assumed them to be.
  Furthermore, the position of these poles are ultraviolet
  divergent.  Thus the standard model in the alpha gauge in
  general, and the Feynman gauge in particular, is not
  renormalizable.  We study also the case with the gauge fixing
  terms (1.3), and find that the propagators remain
  non-renormalizable.  The only gauge without these difficulties
  is the Landau gauge.  As emphasized by Bonneau[1], one
  must make a distinction between the renormalizability of the Green
  functions and that of the physical scattering amplitudes.
\end{abstract}

\newpage

\mysection{Introduction}

In this paper, we shall explore some of the exact and
non-perturbative consequences of the Ward-Takahashi identities in
the standard model.

The organization of this paper will be as follows: in
Sec. 2, we study the $2 \times 2$ mixing matrix of
propagators for the longitudinal~$W$ and the charged Higgs meson
with the gauge fixing term of $W$ chosen to be
$$
\displaystyle
  - \frac{1}{\alpha_W} (\partial^\mu W^+_\mu + i \alpha_W M_0
            \phi^+)
    (\partial^\nu W^-_\nu - i \alpha_W M_0 \phi^-) \, ,
\eqno(1.1)
$$
%                                                                 (1.1)
where $\phi^\pm$ are the charged Higgs fields, $M_0$ is the bare
mass\footnote{%
The bare mass $M_0$ is equal to
$\frac{1}{2} g_0 v_0$, where $v_0$ may be defined either as the
classical vacuum value or the quantum vacuum value of the Higgs
field.  The Green functions differ with different definitions of
$v_0$, but by gauge invariance, physical quantities remain the
same with either definition.
} 
of $W$, and $\alpha_W$ is the
gauge parameter. This study has been carried out by a number of
authors[2,3]. Indeed, we could have started
where they left off, but we opt for making the presentation more
self-contained.

In Sec. 3, we study the $3 \times 3$ mixing matrix of
propagators for the longitudinal~$A$, the longitudinal~$Z$ and
$\phi^0$, where $\phi^0$ is the unphysical neutral Higgs meson,
(the imaginary part of the neutral Higgs meson) with the gauge
fixing term chosen to be
$$
\displaystyle
  - \frac{1}{2 \alpha_Z} (\partial_\mu Z^\mu + \alpha_Z M_0 
  \phi^0)^2 - \frac{1}{2 \alpha_A} (\partial_\mu A^\mu)^2 \, ,
\eqno(1.2)
$$
%                                                                 (1.2)
where $M'_0$ is the bare mass of $Z$, and $\alpha_Z$ and
$\alpha_A$ are gauge parameters.

In Sec. 4, we study these propagators when the gauge
fixing terms are, instead of (1.1) and (1.2),
$$
\displaystyle
  - \frac{1}{2 \alpha_Z} (\partial_\mu Z^\mu)^2
  - \frac{1}{\alpha_W} (\partial^\mu W^+_\mu)
  (\partial^\nu W^-_\nu)
  - \frac{1}{2 \alpha_A}
  (\partial_\mu A^\mu)^2 \, .
\eqno(1.3)
$$
In Sec. 5, we discuss the meaning of our results.

In Appendix A, we list the BRST variations of the
fields in the standard model, in the gauge of (1.1) and
(1.2). In Appendix B, we present an
alternative way to derive the Ward-Takahashi
identities[4], and list the three Ward-Takahashi
identities for propagators involving the longitudinal~$W$, in the
gauge of (1.1). In Appendix C, we list the
nine Ward-Takahashi identities for propagators involving the
longitudinal~$A$ and the longitudinal~$Z$, in the gauge of
(1.2). In Appendix D, we present the
derivations of the three relations satisfied by the 1PI
amplitudes of the propagators in the $3 \times 3$ mixing matrix,
in the gauge of (1.2).  In Appendix E, we
list the twelve Ward-Takahashi identities in the gauge
of (1.3), and derive the relations satisfied by the 1PI
self-energy amplitudes.

\mysection{The Mixing of $W$ and the Charged Higgs Meson}

In the standard model, a longitudinal~$W$ meson may propagate
into either a longitudinal~$W$ or a Higgs meson with the same
charge.  Thus all of the following propagators
$$
\begin{array}{clcr} 
    < 0 | T W^+_\mu (x) W^-_\nu (0) | 0 >  \, , 
       &  < 0 | T W^+_\mu (x) \phi^- (0) | 0 > \, ,  \\
    < 0 | T \phi^+ (x) W^-_\mu (0) | 0 > \, , 
       &  <  0 | T \phi^+ (x) \phi^- (0) | 0 > \, ,      
  \end{array}
$$
are none-zero and together they form a $2 \times 2$ mixing
matrix. We shall denote the Fourier transform of such a
propagator with the symbol $G$ and put
$$
\begin{array}{clcr}
  G^{W^+ W^-} (k) &\equiv&
      -i D^T_{WW} (k^2) T_{\mu \nu} -i \alpha_W D_{WW} (k^2)
      L_{\mu \nu} \, , \\
  G^{W^+ \phi^-}_\mu (k) &=& G^{\phi^+ W^-}_\mu (k)
      \equiv \displaystyle \frac{i k^\mu}{M_0} \alpha_W D_{W \phi} (k^2) \, , 
      \\
  G^{\phi^+ \phi^-} (k) &\equiv& i D_{\phi^+ \phi^-} (k^2) \, , \\
\end{array}
$$
where 
$$
  T_{\mu \nu} \equiv g_{\mu \nu} - \frac{k_\mu k_\nu}{k^2} \, ,
      \qquad L_{\mu \nu} \equiv \frac{k_\mu k_\nu}{k^2} \, .
$$
The $D$ functions are so defined that their unperturbed forms in the
gauge of (1.1) are simply
$$
\displaystyle
    D^{(0)}_{WW} (k^2) = D^{(0)}_{\phi^+ \phi^-} (k^2) 
        = \frac{1}{k^2 - \alpha_w M^2_0}
\eqno(2.1)
$$
They are finite and non-zero at $\alpha_W = 0$.  The unperturbed
form for $D_{W\phi}$ is zero.

The charged Higgs meson mixes not with the transverse $W$ but
with the longitudinal~$W$, as is indicated by the factor $k^\mu$
in the expression for $G^{W^+ \phi^-}$.  Thus, in the $2 \times
2$ mixing matrix under discussion, this factor $k^\mu$ can be
replaced by $K$, where
$$
  K \equiv \sqrt{k^2}
$$
For the same reason, $k^{\mu} k^{\nu} /k^2$ can
be replaced by unity.  The $2 \times 2$ mixing matrix is
therefore equal to
$$
   \left[
    \begin{array}{cc}
      -i \alpha_W D_{WW}  
          &  i K \alpha_W D_{W \phi} / M_0 \\[1ex]
      i K \alpha_W D_{W \phi} / M_0
          & i D_{\phi^+ \phi^-}
    \end{array}
  \right] \, .
\eqno(2.2)
$$
We shall express the propagators by their 1PI amplitudes.

In a field theory without mixing, let the unperturbed propagator
for a particle be $i(k^2-m^2)^{-1}$, and let the 1PI
self-energy amplitude be $\Pi$, then this propagator is the
inverse of $-i (k^2 - m^2 - \Pi)$.  Now that the unperturbed
propagators in the mixing matrix are given by (2.1), the
mixing matrix (2.2) is equal to the inverse of
$$
  \left[
    \begin{array}{cc}
      i \left( k^2 - M^2_0 -\alpha_W  \Pi_{WW} \right) / \alpha_W
          & i K \Pi_{W \phi} / M_0 \\[1ex]
      i K \Pi_{W \phi} / M_0
          & -i (k^2 - \alpha_W M^2_0 - \Pi_{\phi^+ \phi^-})
    \end{array}
  \right] \, ,
\eqno(2.3)
$$
where $\Pi_{W \phi}$, for example, is the 1PI amplitudes for
$D_{W \phi}$. The 1PI amplitudes are functions of $k^2$ and
$\alpha$, the dependence of which will be exhibited only when
needed. Inverting the matrix in (2.3), we find that the
mixing matrix in (2.2) is equal to
\begin{eqnarray}
  &  \left[
    \begin{array}{cc}
      -i (k^2 - \alpha_W M^2_0 - \Pi_{\phi^+ \phi^-}) \alpha_W
          & -i K \alpha_W \Pi_{W \phi} / M_0 \\[1ex]
      -i K \alpha_W \Pi_{W \phi} / M_0
          & i (k^2 - \alpha_W M^2_0 - \alpha_W \Pi_{WW})
    \end{array}
  \right]  &   \nonumber \\[-.5ex]
  &  \rule{4.4in}{1pt}& \, . 
       \label{eq:2.4}  \\[-.5ex]
  &  (k^2 - \alpha_W M^2_0 - \Pi_{\phi^+ \phi^-})
    (k^2 - \alpha_W M^2_0 - \alpha_W \Pi_{WW})
    + k^2 \alpha_W \Pi^2_{W \phi} M^{-2}_0  &
    \nonumber
\end{eqnarray}
Substituting the matrix elements in (2.4) into the Ward
identity (B.4a), we get[2,3] 
$$
  (M^2_0 + \Pi_{WW})
    (M^2_0 - M^2_0 \Pi_{\phi^+ \phi^-} / k^2)
    = (M^2_0 + \Pi_{W \phi})^2 \, ,
\eqno(2.5)
$$
As a side remark, (2.5) provides a subtraction
condition. By setting $k^2$ to zero in (2.5), we find
that
$$
  \Pi_{\phi^+ \phi^-} (0) = 0 \, .
\eqno(2.6)
$$
This ensures the divergence of $\Pi_{\phi^+ \phi^-}$  to be a
logarithmic one. If we express the real part of the neutral Higgs 
field as $(v_0+H)$, and choose $v_0$ to be the quantum value of
the Higgs field, then the Higgs potential has a term linear in
$H$.  This term acts as a counter term which cancels all
contributions of tadpole diagrams.  With the absence of the
contributions of the tadpole diagrams, the self-energy amplitude
$\Pi_{W\Phi}$ is only logarithmically divergent, not
quadratically divergent.  By (2.5), so is $\Pi_{WW}$.
We also note that, with $v_0$ the quantum vacuum value, the Higgs 
potential has a mass term for $\phi^{\pm}$.  When we calculate
$\Pi_{\phi^+ \Phi^-}$, we include the contributions of this mass
term.  But to incorporate the condition (2.6), we make
a subtraction for $\Pi_{\phi^+ \Phi^-}$ at $k^2 = 0$.  The
contributions of the mass term vanish after subtraction.

The denominator in (2.4) is simplified if we take
advantage of (2.5).  We may, for example,
use (2.5) to eliminate $\Pi_{\phi^+ \phi^-}$ from the
denominator, which then becomes
$$
\displaystyle
  \frac{J^2_W}{1 + \Pi_{WW} M^{-2}_0} \, ,
\eqno(2.7)
$$
where
$$
\displaystyle
  J_W \equiv k^2 - \alpha_W M^2_0 - \alpha_W \Pi_{WW} 
     + \frac{k^2}{M^2_0} \Pi_{w \phi} \, .
\eqno(2.8)
$$
Thus we have
$$
\displaystyle
    D_{WW} = \frac{(k^2 - \alpha_W M^2_0 - \Pi_{\phi^+
        \phi^-}) (1 + \Pi_{WW} M^{-2}_0)}
        {(J_W)^2} \, ,
\eqno(2.9a)
$$
$$
\displaystyle
    D_{\phi^+ \phi^-} 
        = \frac{(k^2 - \alpha_W M^2_0 - \alpha_w \Pi_{WW})
            (1 + \Pi_{WW} M^{-2}_0)}
            {(J_W)^2} \, ,
\eqno(2.9b)
$$ 
and
$$
\displaystyle
    D_{W \phi} = - \frac{\Pi_{W \phi} (1 + \Pi_{WW} M^{-2}_0)}
                   {(J_W)^2}
\eqno(2.9c)
$$
Note that the denominator in the expressions in (2.9) is
the square of $J_W$.  Since the poles of these propagators come
from the zeroes of $J_W$, and since $J_W$ is a linear superposition
of 1PI self-energy amplitudes, which are analytic functions of
$k^2$, the order of the poles of the propagators are always even.

Propagators in quantum field theories generally have simple
poles. It is therefore easy to be lulled into believing that
poles of propagators are simple poles. We have found that this is
often not true when two fields mix under the auspices of the
Ward-Takahashi identities. In particular, the charged $\phi$ and
the longitudinal~$W$ have the same unperturbed mass. Thus the
poles of the unperturbed propagators for these fields are both at
$k^2 = \alpha_W M^2_0$.  As the interactions are turned on and
the propagators form a mixing matrix, the positions of the poles
change but the Ward-Takahashi identities force them to remain to
be at the same point. Thus the two simple poles merge to form a
double pole.

The function $J_W$ contains  divergent integrals. Since no subtraction
conditions are available, we expect that the location of the zero
of $J_w$ be ultraviolet divergent. This is confirmed by a
perturbative calculation. We find that
$J_w$ has a zero at
$$
\displaystyle
  k^2 = \alpha_W M^2_0 
        \biggl\{ 1 + \frac{I}{\mu^2_0} 
          + \frac{g_0^2}{8 \pi^2 \epsilon} 
          \left[ \frac{3}{2} 
               - \frac{s^2 \alpha_A}{4} 
               - \frac{\alpha_W}{2}
               - \frac{s^4 \alpha_Z}{4 c^2} 
          \right] \biggr\} \, ,  
\eqno(2.10)
$$
%                                                   (2.10)
where $\mu^2_0$ is the bare mass squared of the physical Higgs and 
is equal to $\lambda_0 v^2_0/2$, $\lambda_0$ and $v_0$ are the bare four
point coupling and the vacuum expectation value of the $\phi$-field 
respectively.   s and c are the sine and cosine of the Weinberg angle.  
$I$ is the contribution from the 
tadpole and $\epsilon = 4 - D$ , 
with $D$ the space-time dimension.   $I$ is given by
$$
\begin{array}{ll}
 I  =  &  \biggl[ \displaystyle \frac{3 \lambda}{2} + \frac{g^2_0}{2} 
                \left( 1 + \frac{1}{2 c^2} \right)   
        \biggr] I_2 \\
    & + \displaystyle \frac{v_0^2}{64 \pi^2 \epsilon} 
        \biggl[ 3 \lambda^2 + \alpha_W \lambda g^2_0 
                + \frac{\alpha_Z \lambda g^2_0}{2 c^2}
                + 3 g^4_0 \left( 1 + \frac{1}{2 c^4} \right) 
        \biggr] \\ \, .
\end{array}
$$  
$I_2$ is the quadratic divergent term and is given by
$$
\displaystyle
  I_2 = \int \frac{d^D p}{(2 \pi)^D} \frac{1}{p^2} \, .
$$
For comparison, the pole of the transverse part of the W is located at
$$
\displaystyle
  k^2 = M^2_0 \biggl\{ 1 + \frac{2 I}{\mu^2_0} 
                       + \frac{g^2_0}{8 \pi^2 \epsilon}
                      \left[ \frac{17}{3} - \frac{3}{4 c^2} 
                             - \frac{\alpha_W}{2}
                             - \frac{\alpha_Z}{2} 
                             \left( c^2 + \frac{1}{2 c^2} \right) 
                      \right] 
              \biggr\} \, .
$$
In (2.10), we keep only the terms which are ultraviolet
divergent and we have not included the contributions of quark or 
lepton loops.  Since the ratio of (2.10) and the transverse mass
pole in the above expression is infinite, the double pole in (2.10) 
cannot be made finite. 

Let the Fourier transform of $< 0 | T i \eta^+ (x) \xi^- (0) |
0 >$ be denoted~as
$$
  G^{\eta^+ \xi^-} \equiv i D_{\eta + \xi^-} \, ,
$$
where $\eta^+$ and $\xi^-$ are hermitian ghost fields associated
with~$W$.  The Ward-Takahashi identity (B.4b) gives
$$
\displaystyle
  D_{\eta^+ \xi^-} = \frac{1 + \Pi_{W \phi} M^{-2}_0}
                      {J_W (1 + g_0 E)} \, ,
\eqno(2.11)
$$
where $E$ is defined in (B.5).  The ghost propagator has
only one factor of $J_W$ in the denominator.

Since the pole of a function cannot be removed by multiplying the
function by a constant, the propagators cannot be made finite by
wavefunction renormalizations. The standard model in the alpha
gauge in general, and the Feynman gauge in particular, is not
renormalizable.

Let us set $\alpha_W = 0$.  In this limit, (2.8) becomes   
$$
  J_W = k^2 (1 + \Pi_{W \phi} M^{-2}_0) \, .
\eqno(2.12)
$$
Thus we have, in the Landau gauge, 
$$
\displaystyle
    D_{\phi^+ \phi^-} = \frac{1}{k^2 - \Pi_{\phi^+ \phi^-}} 
        \, ,  
\eqno(2.13a)
$$ 
and 
$$
\displaystyle  
    D_{\eta^+ \xi^-} = \frac{1}{k^2 (1 + g_0 E)} 
        \, ,   
\eqno(2.13b)
$$  
with the propagators $G^{W^+ \phi^-}$ and $G^{W^+ W^-}_L$ 
vanishing at $\alpha_W = 0$.  Because of (2.6), both
propagators in (2.13) have a simple pole at $k = 0$.
These propagators are renormalizable by wavefunction
renormalizations.

\mysection{The Mixing of $A$, $Z$ and $\phi^0$}

In this section we discuss the $A$, $Z$, and $\phi^0$ mixing with
the gauge fixing terms given by (1.2).

The longitudinal~$Z$ meson mixes with $\phi^0$ as well as with
the longitudinal~$A$, where~$\phi^0$ is the imaginary part of the
neutral Higgs meson and is unphysical.  Thus the propagators for
these fields form a $3 \times 3$ matrix.  Let the propagators be
denoted~as
$$
\begin{array}{clcr}
  G^{AA}_{\mu \nu} (k) &\equiv& 
      -i D^T_{AA} (k^2) T_{\mu \nu} - i \alpha_A D_{AA} (k^2)
      L_{\mu \nu} \, , \\
  G^{ZZ}_{\mu \nu} (k) &\equiv& 
      -i D^T_{ZZ} (k^2) T_{\mu \nu} - i \alpha_Z D_{ZZ} (k^2)
      L_{\mu \nu} \, , \\
  G^{AZ}_{\mu \nu} (k) &=& 
      G^{ZA}_{\mu \nu} (k) \equiv -i D^T_{AZ} (k^2) T_{\mu \nu} 
      - i \alpha_A \alpha_Z D_{AZ} (k^2) L_{\mu \nu} \, , \\
G^{A \phi}_{\mu} (k) &=& - G^{\phi A}_{\mu} (k) 
      \equiv \displaystyle \alpha_A \frac{k \mu}{M'_0} D_{A \phi} (k^2) \, , \\
G^{Z \phi}_{\mu} (k) &=& - G^{\phi Z}_{\mu} (k) 
      \equiv \displaystyle \alpha_Z \frac{k \mu}{M'_0} D_{Z \phi} (k^2) \, , \\
  G^{\phi^0 \phi^0} (k) &\equiv& 
      i D_{\phi^0 \phi^0} (k^2) \, ,
\end{array}
$$
where $M'_0$ is the bare mass of $Z$.  In the above, the function
$G^{AZ}_{\mu \nu}$, for example, is the Fourier transform of
$$
  < 0 | T A_\mu (x) Z_\nu (0) | 0 > \, .
$$
These propagators form the mixing matrix 
$$
  \left[
    \begin{array}{ccc}
      -i \alpha_A D_{AA} & -i \alpha_A \alpha_Z D_{AZ}
                & K \alpha_A D_{A \phi} / M'_0 
                \\[1.5ex] 
      -i \alpha_A \alpha_Z D_{AZ} & -i \alpha_Z D_{ZZ}
                & K \alpha_Z D_{Z \phi} / M'_0
                \\[1.5ex]
      - K \alpha_A D_{A \phi} / M'_0
                & - K \alpha_Z D_{Z \phi} / M'_0
                & i D_{\phi^0 \phi^0}
    \end{array}
  \right]
\eqno(3.1)
$$
The unperturbed forms for these propagators are
$$
\begin{array}{ll}
  D^{(0)}_{AA} (k^2) &= \displaystyle \frac{1}{k^2} \, , \\
  D^{(0)}_{ZZ} (k^2) &= \displaystyle D^{(0)}_{\phi^0 \phi^0} (k^2) 
                     =   \frac{1}{k^2 - \alpha_Z M^{\prime \, 2}_0}
                     \, ,
\end{array}
$$ 
with all other unperturbed forms vanishing. Thus the matrix
in (3.1) is the inverse~of
$$
  \left[
    \begin{array}{ccc}
      i  \left(k^2 - \alpha_A \Pi_{AA} \right) / {\alpha_A}
         & -i \Pi_{AZ}  
         & K \Pi_{A \phi} / M'_0 \\[1.5ex]
      -i \Pi_{AZ} 
         & i \left( k^2 - \alpha_Z M^{\prime \, 2}_0 - \alpha_Z 
           \Pi_{ZZ} \right) / \alpha_Z  
         & K \Pi_{Z \phi} / M'_0 \\[1.5ex]
      - K \Pi_{A \phi} / M'_0 
         & - K \Pi_{Z \phi} / M'_0 
         & -i \left( k^2 - \alpha_Z M^{\prime \, 2}_0
           - \Pi_{\phi^0 \phi^0}  \right)
    \end{array}
  \right]
\eqno(3.2)
$$
where $\Pi_{AA}$, for example, is the 1PI amplitude for $D_{AA}$.
There are nine Ward-Taka\-ha\-shi identities for these propagators,
the derivation of which is given in Appendix C.
Three of these identities give the following three relations
among the 1PI amplitudes,[2,3]:
$$
    (M^{\prime \, 2}_0 + \Pi_{ZZ}) 
        (M^{\prime \, 2}_0 - M^{\prime \, 2}_0 \Pi_{\phi^0 \phi^0} / k^2)
        = (M^{\prime \, 2}_0 + \Pi_{Z \phi})^2 \, ,
\eqno(3.3a)
$$ 
$$
    (M^{\prime \, 2}_0 + \Pi_{ZZ}) \Pi_{AA}
        = \Pi^2_{AZ} \, ,
\eqno(3.3b)
$$
and
$$
    (M^{\prime \, 2}_0 + \Pi_{Z \phi}) \Pi_{AZ}
        = (M^{\prime \, 2}_0 + \Pi_{ZZ}) \Pi_{A \phi} \, .
\eqno(3.3c)
$$ 
The derivation of these relations is presented in
Appendix D.  Note the resemblance of (2.5)
with (3.3a), indeed with (3.3b)
and (3.3c) as well, if one takes into account that the
photon is massless. 

Next we calculate the inverse of (3.2) and equate it to
the matrix in (3.1).  We defer the details to
Appendix D and give only the results here:
$$
\displaystyle 
    D_{AA} (k^2) = \frac{1}{k^2} \, , 
\eqno(3.4a) 
$$
$$
\displaystyle  
    D_{ZZ} (k^2) = 
       \frac{\left[ k^2 (k^2 - \alpha_Z M^{\prime \, 2}_0)
                    + \alpha_A \alpha_Z M^{\prime \, 2}_0 \Pi_{AA}
                    - k^2 \Pi_{\phi^0 \phi^0} \right]
                  \left[ 1 + \frac{\Pi_{ZZ}}{M^{\prime \, 2}_0} \right]}
             {k^2 J^2_Z} \, , 
\eqno(3.4b)
$$
$$
\displaystyle  
    D_{AZ} (k^2) = - D_{A \phi^0} (k^2) 
          = \frac{\Pi_{AZ}}{k^2 J_Z} \, , 
\eqno(3.4c) 
$$
$$
\displaystyle  
    D_{Z \phi} (k^2) = 
       - \frac{\left( \alpha_A M^{\prime \, 2}_0 \Pi_{AA}
                    + \Pi_{Z \phi} k^2 \right)
                  \left( 1 + \frac{\Pi_{ZZ}}{M^{\prime \, 2}_0} \right)}
             {k^2 J^2_Z} \, ,  
\eqno(3.4d)
$$
and
$$
\displaystyle 
    D_{\phi^0 \phi^0} (k^2) = 
        \frac{\left( k^2 - \alpha_Z M^{\prime \, 2}_0 
                    - \alpha_Z \Pi_{ZZ}
                    - \alpha_A \Pi_{AA} \right)
                  \left( 1 + \frac{\Pi_{ZZ}}{M^{\prime \, 2}_0} \right)}
             {J^2_Z} \, ,  
\eqno(3.4e)
$$ 
where 
$$
  J_Z = k^2 - \alpha_Z M^{\prime \, 2}_0 - \alpha_Z \Pi_{ZZ} + k^2
  \frac{\Pi_{Z \phi}}{M^{\prime \, 2}_0} \, .
\eqno(3.5)
$$
Note that all the propagators above except $D_{AA}$ have a double
pole at the simple zero of $J_Z$.  The existence of this double
pole is again easy to understand.  The unperturbed
longitudinal~$Z$ and the unperturbed $\phi^0$ have the same mass,
while the unperturbed $A$ has zero mass.  Thus $D_{ZZ}$ and
$D_{\phi^0 \phi^0}$ have simple poles at the same position, while
$D_{AA}$ has a simple pole at $k^2= 0$.  As the interactions are
turned on, the Ward-Takahashi identities require that the
position of the former two poles remain to be the same, while
that of the last pole remains to be zero.  Thus the propagators
in (3.4) have a double pole as well as a simple pole at
$k^2 = 0$.

A perturbative calculation shows that the
zero of $J_Z$ is located at
$$
\displaystyle
  k^2 = \alpha_Z M^{\prime \, 2}_0 
        \biggl\{ 1 + \frac{I}{\mu^2_0} 
                 + \frac{g^2_0}{8 \pi^2 \epsilon} 
                 \left[ \frac{3 c^2}{2} - \frac{\alpha_Z}{4 c^2}
                   - \frac{s^2}{2} \alpha_W 
                 \right]
        \biggr\} \, . 
\eqno(3.6)
$$
These propagators cannot be made
finite by wavefunction renormalizations.  One can easily check that
this double pole is different from the pole of the transverse
part of the photon-Z mixing sector. 

From the Ward-Takahashi
identities (C.4)--(C.7), we get
$$
\displaystyle
    D_{\eta_A \xi_A} (k^2) = 
       \frac{(1 + g_0 c F_Z) J_Z + \alpha_Z e_0 F_Z \Pi_{AZ}}
       {k^2 J_Z (1 + e_0 F_A + g_0 c F_Z)} \, ,
\eqno(3.7a)
$$
$$
\displaystyle 
    D_{\eta_A \xi_Z} (k^2) = 
       - \frac{\alpha_Z (1 + e_0 F_A) \Pi_{AZ} + g_0 c F_A J_Z}
       {k^2 J_Z (1 + e_0 F_A + g_0 c F_Z)} \, ,    
\eqno(3.7b)
$$
$$
\displaystyle
    D_{\eta_Z \xi_A} (k^2) = -
       \frac{e_0 F_Z (1 + \Pi_{Z \phi} / M^{\prime \, 2}_0)}
       {J_Z (1 + e_0 F_A + g_0 c F_Z)} \, ,    
\eqno(3.7c)
$$
$$
\displaystyle 
    D_{\eta_Z \xi_Z} (k^2) = 
       \frac{(1 + e_0 F_A) (1 + \Pi_{Z \phi} / M^{\prime \, 2}_0)}
       {J_Z (1 + e_0 F_A + g_0 c F_Z)} 
\eqno(3.7d)
$$ 
which express the ghost propagators in terms of the 1PI
self-energy amplitudes as well as the three-point functions $F_A$
and $F_Z$ defined by the equations following (C.4).

Note that the ghost propagators in (3.7) have 
poles at $k^2 = 0$ as well as at a zero of~$J_Z$.  This is because the
unperturbed propagators of the ghosts have a simple pole at $k^2
= 0$ and a simple pole at $k^2 = \alpha_Z M^{\prime \, 2}_0$, same
as the position of the unperturbed $D^{(0)}_{\phi^0 \phi^0}$.  As
interactions are turned on, the Ward-Takahashi identities require
that the former remains to be at $k^2 = 0$, while the latter remains
to be at the same position as the pole of $D_{\phi^0 \phi^0}$.

Finally, we go to the Landau gauge by setting all alphas to zero.
Then the only non-zero $G$ for the unphysical mesons is
$$
\displaystyle
  G^{\phi^0 \phi^0} = \frac{i}{k^2 - \Pi_{\phi^0 \phi^0}} \, ,
\eqno(3.8)
$$
which is logarithmically divergent and is renormalizable by a wave
function renormalization of~$\phi^0$. 

Also, as we set $\alpha_Z$ to zero, we find that
$$
  D_{\eta_A \xi_Z} = D_{\eta_Z \xi_A} \, , 
$$
and the mixing matrix of the propagators of the neutral ghosts is
symmetric.  Such a matrix can be diagonalized by an orthogonal
transformation.  We may therefore renormalize these ghost
propagators by renormalizing the rotated ghost fields obtained by
diagonalization.

\mysection{The Pure Alpha Gauge}

We have shown that, if the alphas are not zero, the propagators
in the preceding two sections have double poles with positions
which are ultraviolet divergent.  Consequently, the standard
model with the gauge fixing terms of (1.1) and
(1.2) are not renormalizable.  We emphasize that the
divergence of the double pole is sufficient but not necessary for 
the theory of the unrenormalizable.  An example is provided by
the quantum theory of the standard model with the gauge fixing
terms those in (1.3).

The unperturbed propagators of the fields in this gauge are given
in (E.4).  We see from the matrix which follows (E.4)
that if we set $M'_0$ to zero, the off-diagonal propagators vanish
while both diagonal propagators have a simple pole at $k^2 = 0$.
In this limit, the charged Higgs meson is a Goldstone boson
decoupled from the longitudinal~$W$, the latter meson being also
massless as a result of the gauge condition.  Next we turn $M$ to
a non-zero value.  The charged Higgs meson remains to be a
Goldstone boson but now it couples with~$W$, which also remains
massless because of the gauge condition.  The two simple poles at
$k^2 = 0$ merge and form a double pole at $k^2 = 0$, (The
 propagator $D_{WW}$ has only a simple pole at $k^2 = 0$
because of the gauge condition.)
Similar considerations hold for the propagators of $A$, $Z$, and $\phi^0$.

To see what happens when the coupling constants are turned on, we
first derive the twelve Ward-Takahashi identities satisfied by
the two-point functions.  These identities are listed
in (E.3).  They are somewhat different in forms from
their counterparts in the gauge of (1.1)
and (1.2), but they lead to the same relations among the
1PI self-energy amplitudes given by (2.5)
and (3.3).

Using (2.5) and (3.3), we get the following
expressions for the propagators:~  %%% kludge to fix bug, don't remove ~
$$
\displaystyle 
    D_{WW} = D_{AA} = D_{ZZ} = \frac{1}{k^2} \, ,
\eqno(4.1a)
$$
$$ 
    D_{AZ} = 0  \, ,
\eqno(4.1b)
$$
$$
\displaystyle  
    D_{\phi^{+} \phi^{-}} = 
       \frac{k^2 - \alpha_W M^2_0 - \alpha_W \Pi_{WW}}
         {(k^2)^2 (1 - \Pi_{\phi^{+} \phi^{-}} / k^2) } \, ,
\eqno(4.1c)
$$
$$
\displaystyle 
    D_{\phi^0 \phi^0} = 
       \frac{k^2 - \alpha_Z M^{\prime \, 2}_0 - \alpha_Z \Pi_{ZZ}
             - \alpha_A \Pi_{AA}}
       {(k^2)^2 (1 - \Pi_{\phi^0 \phi^0} / k^2)} \, ,
\eqno(4.1d)
$$
$$
\displaystyle   
    D_{W \phi} = 
       - \frac{M^{2}_0 + \Pi_{W \phi}}
         {(k^2)^2 (1 - \Pi_{\phi^+ \phi^-} / k^2)} \, ,
\eqno(4.1e)
$$
$$
\displaystyle  
    D_{A \phi} = 
      -  \frac{\Pi_{A \phi}}
         {(k^2)^2 (1 - \Pi_{\phi^0 \phi^0} / k^2)} \, ,
\eqno(4.1f)
$$
$$
\displaystyle  
    D_{Z \phi} = 
       - \frac{M^{\prime \, 2}_0 + \Pi_{Z \phi}}
         {(k^2)^2 (1 - \Pi_{\phi^0 \phi^0} / k^2)} \, .
\eqno(4.1g)
$$ 
We see that the double pole for the $2 \times 2$ mixing matrix
remains located at \hbox{$k^2 = 0$}.  Thus the mixing of the
unphysical Higgs mesons with the longitudinal gauge mesons does
not change the massless nature of these particles as a result of
the Ward identities.

Although the double pole is located at $k^2 = 0$, not infinity,
the propagators still cannot be made finite by renormalizations.
To see this, we first note that (4.1a) says that there
are no radiative corrections to the propagators
in (4.1a), which are already finite without being
divided by wavefunction renormalization constants.  Indeed, if we
were to divide the longitudinal~$A$ or the longitudinal~$Z$ (or
their rotated fields) by wavefunction renormalization constants
which have ultraviolet divergences, the resulting renormalized
propagators for these fields would be ultraviolet divergent.
Thus the longitudinal~$A$ and the longitudinal~$Z$ need no
renormalizations, and the only fields which we may renormalize
are the Higgs fields.  But it is not possible to make the
propagators finite by doing so.  To see this, let the
wavefunction renormalization constant for $\phi^{\pm}$ be
$Z_\phi$.  Then $G^{W \phi}$ multiplied by $\sqrt{Z_\phi}$ and
$G^{\phi^+ \phi^-}$ multiplied by $Z_\phi$ are the renormalized
propagators.  If both renormalized propagators are finite, so is
the ratio $M^2_0 D_{\phi^+ \phi^-} / D^2_{W \phi}$.  But this
ratio is
$$
  (k^2)^2 \left[ \frac{k^2}{M^{\prime \, 2}_0 + \Pi_{WW} (k^2)} 
                 - \alpha_W  \right] \, . 
\eqno(4.2)
$$
%                                                                     (4.2)
Let us examine this expression in the limit of $k^2 \rightarrow
0$.  As we have mentioned, the 1PI self-energy amplitude in this
limit for the longitudinal vector meson is the same as that for
the transverse vector.  Thus the first term inside the bracket
in (4.2) is equal to $k^2$ times the propagator of the
tranverse~$W$ at zero momentum, and is ultraviolet divergent.
Thus the standard model in the gauge of (1.3) is not
renormalizable.  As before, the difficulty of renormalization
disappears as we set all alphas to zero.  Indeed, (4.1c)
and (4.1d) are in the same forms as (2.13a)
and (3.10) as we set all alphas to zero.

Finally, the Ward-Takahashi identities (E.2b),
and (E.3c)--(E.3f) enable us to express the
ghost propagators as
$$
\displaystyle 
    D_{\eta^+ \xi^-} (k^2) =
        \frac{1}{k^2 (1 + g_0 E)} \, ,
\eqno(4.3a)
$$
$$
\displaystyle 
    D_{\eta_A \xi_A} (k^2) =
        \frac{1}{k^2} \frac{1 + g_0 c F_Z}{1 + e_0 F_A + g_0 c F_Z} \, ,
\eqno(4.3b)
$$
$$
\displaystyle  
    D_{\eta_A\xi_Z} (k^2) =
        - \frac{1}{k^2} \frac{g_0 c F_A}{1 + e_0 F_A + g_0 c F_Z} \, ,
\eqno(4.3c)
$$
$$
\displaystyle  
    D_{\eta_Z \xi_Z} (k^2) =
         \frac{1}{k^2} \frac{1 + e_0 F_A}{1 + e_0 F_A + g_0 c F_Z} \, ,
\eqno(4.3d)
$$
$$
\displaystyle  
    D_{\eta_Z\xi_A} (k^2) =
        - \frac{1}{k^2} \frac{e_0 F_Z}{1 + e_0 F_A + g_0 c F_Z} \, .
\eqno(4.3e)
$$
Note that the ghost propagators in (4.3) are identical
in form with the corresponding ones in (3.7) if we set
$\alpha_Z$ to zero, and can be renormalized for the same reason
as before.

\mysection{Conclusion}

Recognizing that the Ward-Takahashi identities provide
constraints on the divergences among amplitudes in the theory,
people have accepted, ever since the appearance of the pioneering
works of t'Hooft and Veltman [5,6], that the BRST
invariance of the standard model guarantees the renormalizability
of the quantum theory of the model in the alpha gauge.

In the standard model, there are twelve Ward-Takahashi identities
for the two-point functions.  Four of these identities lead to
four relations satisfied by the nine 1PI self-energy amplitudes
of the unphysical mesons, leaving five of the 1PI self-energy
amplitudes independent.  Five other Ward-Takahashi identities
relate the five independent ghost propagators and three 3-point
functions to these same 1PI self-energy amplitudes.  The last
three Ward-Takahashi identities set the mass scale of the
physical vector mesons.  For example, the Ward-Takahashi
identity (B.6) leads to a relation between the $W$-mass
and the quantum vacuum expectation of the Higgs meson, as will be
discussed in more details in another paper.  We shall only point
out here that, in the gauge of (1.1) and (1.2),
with no conditions of subtraction available for the position of
the double pole in the unphysical propagators, there is nothing
to prevent it from being infinite.

Over a year ago, we first realized that, in the Abelian gauge
field theory with Higgs mesons, with a gauge fixing term of the
form of (1.2), one of the Ward-Takahashi identities
enforces the isolated singularity of the propagators in the $2
\times 2$ mixing matrix in this theory to be a double pole.
Explicit perturbative calculations on the position of this double
pole verified that the position of this pole is ultraviolet
divergent.  Thus the Abelian gauge field theory with Higgs mesons
is not renormalizable in the alpha gauge.

Since then we have shown that all these are also true in the
standard model, both for the $2 \times 2$ mixing matrix and for
the $3 \times 3$ mixing matrix for $v_0$ the classical vacuum
value or the quantum vacuum value.  We have also extended the
treatment to the alpha gauge with the gauge fixing terms
of (1.3).  While the double pole in this case is at $k^2
= 0$, not infinity, the propagators cannot be rendered finite
with wavefunction renormalizations.

There have been many proofs [7,8,9,10] 
demonstrating that gauge field theories with symmetry breaking in
general, and the standard model in particular, are renormalizable
in the alpha gauge.  Instead of addressing everyone of them in
details, we would like to make the following comment.  The
considerations of renormalizability of quantum gauge field
theories with symmetry breaking differ from those without
symmetry breaking.  In the former theories, the unphysical Higgs
mesons mix with the longitudinal component of the associated
gauge meson.  It is untenable to argue that since the latter
theories are renormalizable, so are the former theories.

To illustrate this point, let us consider the simple example of
scalar QED in which the photon couples with a complex scalar
field~$\phi$ which has a real and positive bare mass $\mu_0$, and
the vacuum symmetry is not broken.  To quantize this theory, let
us choose the gauge fixing term
$$
\displaystyle
  - \frac{1}{2 \alpha} (\partial_\mu A^\mu - \alpha M \phi_2)^2
\eqno(5.1)
$$ 
where $\phi_2$ is the imaginary part of $\phi$ and $M$ is an
introduced parameter.  The gauge fixing term for scalar QED is
traditionally chosen to be the one in (5.1) with $M$
equal to zero, but there is nothing to forbid us from choosing a
non-zero $M$ provided that we add the corresponding ghost terms
to make the effective Lagrangian invariant under BRST variations.
While the Green functions for a non-zero $M$ are different from
those with $M$ equal to zero, the physical scattering amplitudes
remain the same.

The gauge fixing term in (5.1) contains a term which
mixes the longitudinal~$A$ with $\phi$.  This changes the tree
amplitudes but not any of the 1PI amplitudes of the propagators.
Thus we have
$$
  \Pi_{AA} = \Pi_{A \phi_2} = 0
$$
and the $2 \times 2$ mixing matrix for the propagators of~$A$ and
$\phi_2$ is equal to the inverse~of
$$
  \left[
  \begin{array}{cc}
     {i k^2 / \alpha}   &    K M    \\
    - K M         &   -i (k^2 - \mu^2_0 - \alpha M^2 
                               - \Pi_{\phi \phi})
  \end{array}
  \right] \, .
$$
Thus this mixing matrix is equal to    
\begin{eqnarray}
  & \left[
    \begin{array}{cc}
    - i \alpha (k^2 - \mu^2_0 - \alpha M^2 - \Pi_{\phi \phi} )  
      &   - \alpha K M    \\[1.5ex]
    \alpha K M     &   i k^2
    \end{array}
  \right]  &   \nonumber \\[-.5ex]
  & \rule{3in}{1pt} & \, . 
  \label{eq:5.2} \\[-.5ex]
  &  k^2 (k^2 - \mu^2_0 -\Pi_{\phi \phi}) &    \nonumber
\end{eqnarray}
%                                                                  (5.2)
We see from (5.2) that the propagators are not
renormalizable for a finite and non-zero~$M$.  In order that the
theory is renormalizable, a non-zero $M$ must be equal to a
finite number times $1 / \sqrt{Z_{\phi}}$, where $Z_{\phi}$ is the
wavefunction renormalization constant for the $\phi_2$-field.

In the standard model there is no such freedom of choosing~$M$,
and the unphysical propagators are not renormalizable in the alpha gauge.
Since graphs of unphysical propagators may appear as subgraphs in
the graphs of other Green functions, Green functions in the
standard model are generally not renormalizable.

The only gauge in which the difficulty of renormalization does
not appear is the Landau gauge, which is obtained from the alpha
gauge by setting alpha to zero.  One of the reasons for this is
that, as all alpha are set to zero, many propagators vanish and
need no renormalization.

This does not necessarily mean that the standard model is
renormalizable in Landau gauge.  But if it is, and if physical
(and on-shell) scattering amplitudes are gauge invariant, these
amplitudes will be finite in the alpha gauge once they are finite 
in the Landau gauge.  In a practical calculation of physical
scattering amplitudes, the infinities from the Green functions
cancel one another, provided that they are properly regularized
and the procedures of subtractions and normalization as
emphasized by Bonneau [11] are properly performed.  On the other hand,
off-shell Green functions are dependent on alpha, and are not renormalizable in the
alpha gauge.  In a gauge field theory with symmetry breaking, one
must make a distinction between the renormalizability of the
Green functions and that of the physical scattering amplitudes.

\newpage
\begin{center}
\large {\bf References}
\end{center}

\begin{enumerate}

\item{  G.~Bonneau, Nucl. Phys. \underline{B221} (1983)~178. }

\item{ L.~Baulieu and R.~Coquereaux, Ann. Phys.
  \underline{140} (1982)~163. }

\item{ See also K.~Aoki, Z.~Hioki, R.~Kawabe, M.~Konuma,
  and T.~Muta, Prog. Theor. Phys. Suppl. \underline{73} 
  (1982)~1, and the references quoted in this paper.  A number of 
  our results, e.g., (2.5), disagree with their
  counterparts in the paper of reference [2], but agree
  with their counterparts in this paper. }

\item{  E.C.~Tsai, (private communication, 1986).  }

\item{ G.~t'Hooft, Nucl. Phys. \underline{B33}
  (1971)~173, ibid. \underline{B35} (1971)~167.  } 

\item{  G.~t'Hooft and M.~Veltman, Nucl. Phys. \underline{B44} (1972)~189. }

\item{  B.W.~Lee and J.~Zinn-Justin, Phys. Rev. \underline{D5}
  (1972) 3121, 3137, 3155.  }

\item{  J.C.~Taylor, Nucl. Phys. \underline{B33} (1971)~436. }

\item{  A.A.~Slavnov, Theor. and Math. Phys. \underline{10} (1972)~99.}

\item{ K.~Fujikawa, B.W.~Lee, and A.I.~Sanda,
  Phys. Rev. \underline{D6} (1972)~2923.  } 

\item{  G.~Bonneau, Int. J Mod. Phys. \underline{5} (1990)~3831. }

\end{enumerate}

\newpage

\begin{center}
\large \bf {Appendix A}
\end{center}

We list some of the BRST variations of fields in the standard
model below:
$$
  \delta W^\pm_\mu = 
               \partial_{\mu} \xi^{\pm} \pm ig_0
               (c Z_{\mu} + s A_{\mu}) \xi^{\pm} \mp
               ig_0 W^{\pm}_{\mu} (c \xi_Z + s \xi_A) \, ,
\eqno(A.1)
$$
$$ 
      \delta Z_\mu = 
              \partial_{\mu} \xi_Z + ig_0 c
              ( W^+_{\mu} \xi^- -W^-_{\mu} \xi^+) \, ,
\eqno(A.2)
$$
$$ 
    \delta A_\mu = 
              \partial_{\mu} \xi_A +ig_0
              s (W^+_{\mu} \xi^- -W^-_{\mu} \xi^+) \, ,
\eqno(A.3)
$$
$$
\displaystyle
     \delta \phi^\pm = 
             \mp ig_0 \left[ \left( \frac{c^2-s^2}{2c}
                 \xi_Z + s \xi_A \right)
                 \phi^{\pm} + \frac{1}{2} \xi^{\pm}
                 (v_0 + H \pm i \phi^0) \right] \, ,
\eqno(A.4)
$$
$$
\displaystyle
     \delta H =
              - \frac{g_0}{2c} \xi_Z \phi 
              - \frac{ig_0}{2} 
              (\xi^- \phi^+ - \xi^+ \phi^-) \, ,
\eqno(A.5)
$$
$$
\displaystyle
  \delta \phi^0 = \frac{g_0}{2c} \xi_Z (v_0 +H)
             - \frac{g_0}{2} (\xi^- \phi^+ + \xi^+ \phi^-) \, ,
\eqno(A.6)
$$
$$
\displaystyle
  \delta i \eta^\pm =
           \frac{1}{\alpha_W} (\partial^{\mu} W^{\pm}_{\mu} \pm
           i \alpha_W M_0 \phi^{\pm}) \, ,
\eqno(A.7)
$$
$$
\displaystyle
  \delta i \eta_Z =
        \frac{1}{\alpha_Z} (\partial_{\mu} Z^{\mu}
        + \alpha M'_0 \phi^0) \, ,
\eqno(A.8)
$$
$$
\displaystyle
  \delta i \eta_A =
            \frac{1}{\alpha_A} \partial_{\mu} A^{\mu} \, .
\eqno(A.9)
$$
In the above,
\begin{eqnarray*}
  W^\pm &\equiv & \frac{W^1_\mu \mp i W^2_\mu}{\sqrt{2}} \, ,
     \\[1ex]
  M_0 &\equiv& \frac{1}{2} g_0 v_0 \, , \\[1ex]
  M'_0 &\equiv& \frac{1}{c} M_0 \, ,
\end{eqnarray*}
$\eta$ and $\xi$ are the hermitian ghost fields, $g_0$ is the bare weak
coupling constant, and $\alpha$ is the gauge parameter.  (We
denote the gauge parameter for~$A$ as $\alpha_A$.)  Also, the
Higgs field is given by
\begin{displaymath}
  \phi \equiv \left( 
    \begin{array}{c}
      \phi^+ \\[1ex]
      (v_0 + H + i \phi^0) / \sqrt{2} 
    \end{array}
\right)
\end{displaymath}

\begin{center}
\large \bf {Appendix B}
\end{center}

In a gauge field theory with an effective Lagrangian satisfying
the BRST invariance, the vacuum state satisfies
$$
  Q | 0 > = 0 \, .
\eqno(B.1)
$$              
In (B.1), $Q$ is the BRST charge the commutator
(anti-commutator) of which with a Bose (Grassmann) field is the
BRST variation of the field.

The Ward-Takahashi identities can be derived directly
from (B.1). We have, as a result of (B.1),
$$
  < 0 | O Q | 0 > = 0 \, ,  
\eqno(B.2)
$$
where $O$ is any operator.  Next we move $Q$ in (B.2)
to the left.  Since
$$
  < 0 | Q = 0 \, ,
$$
we get 
$$
  < 0 | \delta O | 0 > = 0 \, , 
\eqno(B.3)
$$
where $\delta O$ is the BRST variation of~$O$.

Let us next derive the Ward-Takahashi identities for the propagators 
in the mixing matrix for the longitudinal~$W$ and the charged~$\phi$.

By choosing $T (i \eta^+ \delta \eta^-)$ as $O$
in (B.3), where $T$ is the time-ordering operator, we
get
$$
\left< 0 | T \left( \frac{1}{\alpha_W} \partial^\mu W^+_\mu + i M_0
        \phi^+ \right)
      \left( \frac{1}{\alpha_W} \partial^\nu W^-_\nu - i M_0 \phi^- \right)
      | 0 \right> 
    = 0 
$$ 
or
$$
\displaystyle
    \frac{1}{\alpha_W} - \frac{k^2}{\alpha_W} D_{WW} (k^2) 
       - 2 k^2 D_{W \phi} (k^2) + M^2_0 D_{\phi^+ \phi^-} (k^2) 
    = 0 \, .
\eqno(B.4a)
$$
The Ward-Takahashi identity (B.4) relates the
propagators in the $2 \times 2$ mixing matrix.

From $< 0 | \delta T i \eta^+ W^-_\nu | 0 > = 0$, we get
$$
\begin{array}{ll}
\displaystyle
     \left< 0 | T \left( \frac{1}{\alpha_W} \partial^\mu W^+_\mu + i
         M_0 \phi^+ \right)
        W^-_\nu | 0 \right>
       \\
    {\hspace{4.5em}} - \left< 0 | T i \eta^+ 
       ( \partial_\nu \xi^- - i g_0 (c Z_\nu + s 
      A_\nu) \xi^- + i g_0 W^-_\nu (c \xi_Z + s \xi_A)) | 0 \right> = 0 \, ,
\end{array}
$$
which leads to 
$$
  - D_{WW} (k^2) - \alpha_W D_{W \phi} (k^2)
  + D_{\eta^+ \xi^-} (k^2) (1 + g_0 E) = 0
\eqno(B.4b)
$$
where the Fourier transform of $i < 0 |T i \eta^+ [ (c Z_\nu + s
A_\nu) \xi^-  -  W^-_\nu (c \xi_Z + s \xi_A)] | 0 >$ is defined
to be
$$
  D_{\eta^+ \xi^-} (k^2) E k^\nu \, .
\eqno(B.5)
$$
The identity (B.5) relates the ghost propagator $D_{\eta^+
  \xi^- }$ to the propagators in the $2 \times 2$ mixing matrix.

From $< 0 | \delta T i \eta^+ \phi^- | 0 > = 0$, we get
$$
\begin{array}{ll}
  \lefteqn{\hspace{-0.5em}
    \left< 0 | T \left( \frac{1}{\alpha_W} \partial^\mu W^+_\mu + i M_0
        \phi^+ \right)
      \phi^- | 0 \right> 
    } \\
  \hspace{5em} - i g_0 \left< 0 | T i \eta^+ 
    \left[ \left( \displaystyle \frac{c^2 - s^2}{2c} \xi_Z 
        + s \xi_A \right) \phi^- 
      + \frac{1}{2} \xi^- (v_0 + H - i \phi^0) \right] | 0 \right> = 0 \, ,
\end{array}
$$
which leads to 
$$
\displaystyle  
  \frac{k^2}{M_0} D_{W \phi} (k^2) - M_0 D_{\phi^+ \phi^-} (k^2)
  + g_0 D_{\eta^+ \xi^-} (k^2) \left( \frac{< 0 | v_0 + H | 0
      >}{2} + F \right) = 0 \, ,
\eqno(B.6)
$$ 
where the Fourier transform of the connected part of
$$
\displaystyle
  -i \left< 0 | T i \eta^+ \left[ \left( \frac{c^2- s^2}{2c} \xi_Z
      + s \xi_A \right) \phi^- 
      + \frac{1}{2} \xi^- (H - i \phi^0) \right] | 0 \right> 
$$
is defined to be 
$$
  D_{\eta^+ \xi^-} (k^2) F \, .
\eqno(B.7)
$$
The identity (B.6) relates the propagators with the
vacuum expectation value of the Higgs field.

\begin{center}
\large \bf {Appendix C}
\end{center}

In this Appendix, we derive the Ward-Takahashi identities
involving the propagators in the $3 \times 3$ mixing matrix for
the longitudinal~$A$, the longitudinal~$Z$, and $\phi^0$,

From $< 0 | \delta T i \eta_A \partial_\nu A^\nu | 0 > = 0$, we get
$$
\displaystyle
  \frac{1}{\alpha_A} < 0 | T \partial_\mu A^\mu \partial_\nu
  A^\nu | 0 > = 0 \, ,
$$
which leads to
$$
\displaystyle
  D_{AA} (k^2) = \frac{1}{k^2} \, .
\eqno(C.1)
$$
From $\displaystyle < 0 | \delta T i \eta_Z \left( \frac{1}{\alpha_Z}
  \partial_\nu Z^\nu + M'_0 \phi^0 \right) | 0 > = 0$, we get
\begin{displaymath}
  < 0 | T \left( \frac{1}{\partial_Z} \partial_\mu Z^\mu +
    M'_0 \phi^0 \right)
  \left( \frac{1}{\alpha_Z} \partial_\nu Z^\nu + M'_0 \phi^0 \right)
  | 0 > = 0 \, ,
\end{displaymath}
which leads to
$$
\displaystyle
  \frac{1}{\alpha_z} - \frac{k^2}{\alpha_Z} D_{ZZ} 
  - 2 k^2 D_{Z \phi} + M^{\prime \, 2}_0 D_{\phi^0 \phi^0} = 0 \, .
\eqno(C.2)
$$
From $\displaystyle < 0 | \delta T i \eta_A \left( \frac{1}{\alpha_Z}
  \partial_\nu Z^\nu + M'_0 \phi^0 \right) | 0 > = 0$, we get
$$
\displaystyle
  < 0 | T \frac{ \partial_\mu A^\mu}{\alpha_A}
  \left( \frac{1}{\alpha_Z} \partial_\nu Z^\nu 
    + M'_0 \phi^0 \right) | 0 > = 0
$$
which leads to 
$$
  D_{AZ} = - D_{A \phi} \, .
\eqno(C.3)
$$
The three Ward-Takahashi identities derived above are independent
of the propagators for the ghost fields.

Next, from $< 0 | \delta T i \eta_A A_\nu | 0 > = 0$, we get
$$
  \frac{1}{\alpha_A} \left< 0 | T \partial^\mu A_\mu A_\nu | 0 \right>
  - \left< 0 | T i \eta_A \left[ \partial_\nu \xi_A + i e_0
    (W^+_\nu \xi^- - W^-_\nu \xi^+) \right] | 0 \right> = 0 \, ,
$$
which leads to 
$$
\displaystyle
  - \frac{1}{k^2} + D_{\eta_A \xi_A} (k^2) (1 + e_0 F_A)
  + D_{\eta_A \xi_Z} (k^2) e_0 F_Z = 0 \, ,
\eqno(C.4)
$$
where $e_0$ is the bare electric charge and where
\begin{eqnarray*}
  \Gamma_{\eta_A W^+_{\nu} \xi^-} 
  - \Gamma_{\eta_A W^-_{\nu} \xi^+}
  &\equiv & k_{\nu} F_A (k^2) \, , \\
  \Gamma_{\eta_Z W^+_{\nu} \xi^-}
  - \Gamma_{\eta_Z W^-_{\nu} \xi^+} 
  & \equiv & k_{\nu} F_Z (k^2) \, .
\end{eqnarray*}
In the above $\Gamma_{\eta_A W^+_{\nu} \xi^-}$ is the
truncated 3-point function with the fields $W^+_\nu$ and $\xi^-$
joint at the same space-time point.

From $< 0 | \delta T i \eta_A Z_\nu | 0 > = 0$, we get
$$
\displaystyle
  \frac{1}{\alpha_A} \left< 0 | T \partial^\mu A_\mu Z_\nu | 0 \right> 
  - \left< 0 | T i \eta_A \left[ \partial_\nu \xi_Z + i g_0 c
    (W^+_\nu \xi^- - W^-_\nu \xi^+) \right] | 0 \right> = 0
$$
which leads to 
$$
  - \alpha_Z D_{AZ} (k^2) + D_{\eta_A \xi_Z} (k^2) (1 + g_0 c F_Z)
  + D_{\eta_A \xi_A} (k^2) g_0 c F_A = 0 \, .
\eqno(C.5)
$$
From $< 0 | \delta T i \eta_Z A_\nu | 0 > = 0$, we get
$$
  \left< 0 | T \left( \frac{1}{\alpha_Z} \partial^\mu Z_\mu + M'_0
    \phi^0 \right) A_\nu | 0 \right> 
  - \left< 0 | T i \eta_Z \left[ \partial_\nu \xi_A + i e_0 
    (W^+_\nu \xi^- - W^-_\nu \xi^+) \right] | 0 \right> = 0  \, ,
$$
which leads to
$$
  D_{\eta_Z \xi_A} (k^2) (1 + e_0 F_A) 
  + e_0 D_{\eta_Z \xi_Z} (k^2) F_Z = 0 \, .
\eqno(C.6)
$$
From $< 0 | \delta T i \eta_Z Z_\nu | 0 > = 0$, we get
$$
  \left< 0 | T \left( \frac{1}{\alpha_Z} \partial^\mu Z_\mu + M'_0
    \phi^0 \right) Z_\nu | 0 \right> 
  - \left< 0 | T i \eta_Z \left[ \partial_\nu \xi_Z + i g_0 c
    (W^+_\nu \xi^- - W^-_\nu \xi^+) \right] | 0 \right> = 0
$$
which leads to 
$$
  - D_{ZZ} (k^2) - \alpha_Z D_{Z \phi} (k^2) 
  + D_{\eta_Z \xi_Z} (k^2) (1 + g_0 c F_Z) 
  + g_0 c D_{\eta_Z \xi_A} (k^2) F_A = 0 \, .
\eqno(C.7)
$$
The four Ward-Takahashi
identities (C.4)--(C.7) relate the ghost
propagators to the propagators in the mixing matrix.

From $< 0 | \delta T i \eta_A \phi^0 | 0 > = 0$, we get
$$
  \left< 0 | T \frac{1}{\alpha_A} \partial^\mu A_\mu \phi^0 | 0 \right> 
  - \left< 0 | T i \eta_A \left[ M'_0 \xi_Z + \frac{g_0}{2 c} \xi_Z
    H - \frac{g_0}{2} (\xi^- \phi^+ + \xi^+ \phi^-) \right] | 0
  \right> = 0   
$$
which leads to
$$
\begin{array}{ll}
  {\hspace{-3em}
    \displaystyle - \frac{k^2}{M'_0} D_{A \phi} (k^2) 
  - \frac{g_0}{2} D_{\eta_A \xi_Z} (k^2)
  \left[ \frac{< 0 | v_0 + H | 0 >}{c} + \frac{1}{c}
    \Gamma_{\eta_Z\xi_Z H} - \Gamma_{\eta_Z \xi^- \phi^+} -
    \Gamma_{\eta_Z \xi^+ \phi^-} \right] 
      } \\
  {\hspace{6em}} \displaystyle - \frac{g_0}{2} D_{\eta_A \xi_A} (k^2) \left[ \frac{1}{c}
    \Gamma_{\eta_A\xi_Z H} - \Gamma_{\eta_A\xi^-\phi^+}
    - \Gamma_{\eta_A\xi^+\phi^-} \right] = 0 \, .
\end{array}
\eqno(C.8)
$$
where $\Gamma_{\eta_Z \, \xi_Z H}$, say, is the truncated $3$-point
Green function of the fields $\eta_Z$, $\xi_Z$ and $H$, with the
latter two fields joined at the same space-time point.

From $< 0 | \delta T i \eta_Z \phi^0 | 0 > = 0$, we get
\begin{eqnarray*}
  \lefteqn{\hspace{-3em}
    \left< 0 | T \left( \frac{1}{\alpha_Z} \partial^\mu Z_\mu +
      M'_0 \phi^0 \right) \phi^0 | 0 \right> 
     }  \\
  &&{\hspace{2em}} - 
    \left< 0 | T i \eta_Z \left[ M'_0 \xi_Z + \frac{g_0}{2c}
      \xi_Z H - \frac{g_0}{2} \left( \xi^- \phi^+ + \xi^+ \phi^-
      \right) \right] | 0 \right> = 0 \, ,
\end{eqnarray*}
which leads to
$$ 
\begin{array}{ll}
  \lefteqn{ \hspace{-3em}
    {}- \frac{k^2}{M'_0} D_{Z \phi} (k^2) + M'_0 D_{\phi^0 \phi^0}
    (k^2) 
    }   \\
  \lefteqn{ \hspace{-3em}
    {}- \frac{g_0}{2} D_{\eta_Z \xi_Z} (k^2)
      \left[ \frac{< 0 | v_0 + H | 0 >}{c}
        + \frac{1}{c} \Gamma_{\eta_Z\,  \xi_Z H }
        - \Gamma_{\eta_Z\,  \xi^- \phi^+}
        - \Gamma_{\eta_Z \, \xi^+ \phi^- }\right] 
     }  \\
  {\hspace{4em}} \displaystyle - \frac{g_0}{2} D_{\eta_Z \xi_A} (k^2)
  \left[ \frac{1}{c} \Gamma_{\eta_A \, \xi_Z H} - \Gamma_{\eta_A
      \, \xi^-
    \phi^+} 
    - \Gamma_{\eta_A\,  \xi^+ \phi^-} \right] = 0 \, .
\end{array}
\eqno(C.9)
$$
The Ward-Takahashi identities (C.8) and (C.9)
relate the propagators with the vacuum expectation value of the
Higgs field.

\begin{center}
\large \bf {Appendix D}
\end{center}

Equating the matrix in (3.1) to the inverse of the
matrix in (3.2), we express the propagators
in (3.1) by their 1PI amplitudes.  With the propagators
expressed in such forms, we require them to satisfy the
Ward-Takahashi identity (C.3).  We get
$$
  \Pi_{AZ}(k^2-\alpha_Z M^{\prime \, 2}_0
  -\Pi_{\phi^0 \phi^0} -\alpha_Z \Pi_{Z \phi})
  = \Pi_{A \phi} (k^2-\alpha_Z M^{\prime \, 2}_0
  - \alpha_Z \Pi_{ZZ} + k^2 \Pi_{Z \phi}/M^{\prime \, 2}_0) \, .
\eqno(D.1)
$$
Similarly, by requiring (C.1) be satisfied, we get
$$
  \Pi_{AA} (k^2-\alpha_Z M^{\prime \, 2}_0
     - \Pi_{\phi^0 \phi^0} - \alpha_Z \Pi_{Z \phi})
     = \Pi_{A \phi} (-\alpha_Z \Pi_{AZ} +
     k^2 \Pi_{A \phi}/M^{\prime \, 2}) \, ,
\eqno(D.2)
$$
where we have made use of (D.1).  Also,
imposing (C.2) gives,
$$
\displaystyle
  (1+ \Pi_{ZZ}/M^{\prime \, 2}_0)
  (1- \frac{\Pi_{\phi^0 \phi^0}}{k^2})
  =(1+ \Pi_{Z \phi}/M^{\prime \, 2}_0)^2 \, ,
\eqno(D.3)
$$
where we have made use of (D.1) and (D.2) to
eliminate $\Pi_{AA}$ and $\Pi_{A \phi^0}$.  We may
reduce (D.1) further by making use of (D.3) to
eliminate $\Pi_{\phi^0 \phi^0}$ and get
$$
     \Pi_{AZ}(1+ \Pi_{Z \phi^0}/M^{\prime \, 2}_0)
     = \Pi_{A \phi} (1+ \Pi_{ZZ}/M^{\prime \, 2}_0) \, .
\eqno(D.4)
$$
Similarly, we may eliminate $\Pi_{\phi \phi}$ from (D.2)
and get
$$
  (1+ \Pi_{ZZ}/M^{\prime \, 2}_0) (\Pi_{AA}/M^{\prime \, 2}_0)
  = (\Pi_{AZ}/M^{\prime \, 2}_0)^2 \, .
\eqno(D.5)
$$
Note the resemblance of (D.3) and (D.5)
with (2.5).

With (D.3)--(D.5), we may reduce the
determinant of the $3 \times 3$ matrix in (3.2) into
$$
  ik^2 J^2_Z/[\alpha_A \alpha_Z (1+ \Pi_{ZZ}/M^{\prime \, 2}_0)]
\eqno(D.6)
$$
where  
$$
\displaystyle
  J_Z=k^2 - \alpha_Z M^{\prime \, 2}_0 
  -\alpha_Z \Pi_{ZZ} + k^2 \frac{\Pi_{Z\phi}}{M^{\prime \, 2}_0} \, .
\eqno(D.7)
$$

Equations (D.6) and (D.7) are obtained by
eliminating $\Pi_{AA}$ and $\Pi_{A \phi^0}$ from the expression.
Note the similarity between (D.7) and (2.8).

\begin{center}
\large \bf {Appendix E}
\end{center}

In this Appendix, we shall study the propagators in the pure
alpha gauge defined by the gauge fixing terms (1.3).  As
in the preceding appendices, we shall use the Ward-Takahashi
identities in this gauge to determine relations among the 1PI
self-energy amplitudes. We then simplify the expressions for
these propagators by the use of these relations.

In the pure alpha gauge, the BRST variations of the fields remain
the same as the ones given in Appendix A with the
following exceptions
\begin{eqnarray}
  \delta i \eta^+ &=& \frac{1}{\alpha_W} \partial^{\mu}
  W^+_{\mu} \, ,  \nonumber \\
  \delta i \eta_Z
  &=& \frac{1}{\alpha_Z} \partial \delta^{\mu} Z_{\mu} \, . 
     \label{eq:e.1}
\end{eqnarray}

The Ward-Takahashi identities for the propagators of the
longitudinal~$W$ and the unphysical charged $\phi$ give
$$
\displaystyle
    D_{WW} (k^2) = \frac{1}{k^2} \, ,  
\eqno(E.2a)
$$
$$ 
    D_{WW} (k^2) = D_{\eta^+ \xi^-} (k^2) (1 + g_0 E) \, ,
\eqno(E.2b)
$$
where $E$ is given by (B.6) and 
$$
\displaystyle
    \frac{k^2}{M^2_0} D_{W \phi} (k^2)
    = - D_{\eta^+ \xi^-} (k^2)
      \left( Z+ \frac{F}{M_0} g_0 \right) \, ,
\eqno(E.2c)
$$
where $F$ is given by (B.7) and
$$
  Z \equiv < 0| v_0 +H |0 > /v_0 \, .
$$
The Ward-Takahashi identities for the propagators of the
longitudinal~$A$, the longitudinal~$Z$, and the unphysical
neutral $\phi^0$ are
$$
\displaystyle 
    D_{AA} (k^2) = D_{ZZ}(k^2)=\frac{1}{k^2} \, ,
\eqno(E.3a)
$$
$$
    D_{AZ} (k^2) = 0 \, ,
\eqno(E.3b)
$$
$$
\displaystyle 
    - \frac{1}{k^2} + D_{\eta_A \xi_A} (k^2)
       (1+e_0F_A)+ D_{\eta_A \xi_Z} (k^2)
       e_0 F_Z =0 \, ,
\eqno(E.3c)
$$ 
  where $F_A$ and $F_Z$ are defined in the equations
  following (C.4),
$$
    D_{\eta_A \xi_Z} (k^2) (1+g_0 cF_Z) +
    D_{\eta_A \xi_A} (k^2) g_0 c F_A
       = 0 \, , 
\eqno(E.3d)
$$ 
$$
       D_{\eta_Z \xi_A}(k^2)(1+e_0F_A)+
       D_{\eta_Z \xi_Z} (k^2) e_0 F_Z
       = 0 \, , 
\eqno(E.3e)
$$  
$$
\displaystyle 
       - \frac{1}{k^2}+ D_{\eta_Z \xi_Z}(k^2)
       (1+ g_0 c F_Z) + D_{\eta_Z \xi_A}(k^2)
       g_0 c F_A
       = 0 \, , 
\eqno(E.3f)
$$ 
$$ 
\begin{array}{ll}
    \lefteqn{\hspace{-3em}
      - \frac{k^2}{M'_0} D_{A \phi} (k^2)
      - \frac{g_0}{2} D_{\eta_A \xi_Z} (k^2)
      \left[ \frac{ < 0 | v_0 +H |0 >}{c}
        + \frac{1}{c} \Gamma_{\eta_Z\,  \xi_Z H}
        - \Gamma_{\eta_Z\, \xi^- \phi^+}
        - \Gamma_{\eta_Z\,  \xi^+ \phi^-}
      \right]}  \\[1ex]
    {\hspace{3em}} 
       \displaystyle -\frac{g_0}{2} D_{\eta_A \xi_A} (k^2)
       \left[ \frac{1}{c} \Gamma_{\eta_A\,  \xi_Z H}
         - \Gamma_{\eta_A\,  \xi^- \phi^+}
         - \Gamma_{\eta_A \, \xi^+\phi^-} \right]= 0 \, , \\
\end{array}
\eqno(E.3g)
$$ 
and               
$$
\begin{array}{ll}
    \lefteqn{\hspace{-3em}
      - \frac{k^2}{M'_0} D_{Z \phi} (k^2)
      - \frac{g_0}{2} D_{\eta_Z \xi_Z} (k^2)
      \left[ \frac{v_0Z}{c} + \frac{1}{c}
        \Gamma_{\eta_Z\, \xi_ZH}
        - \Gamma_{\eta_Z \,\xi^- \phi^+}
        \Gamma_{\eta_Z \,\xi^+ \phi^-} \right]
      }  \\[1ex]
    {\hspace{4em}} \displaystyle - \frac{g_0}{2}
       D_{\eta_Z \xi_A} (k^2)
       \left[ \frac{1}{c} \Gamma_{\eta_A\, \xi^- H}
           - \Gamma_{\eta_A \, \xi^- \phi^+}
           - \Gamma_{\eta_A \, \xi^+ \phi^- }\right]
       = 0 \, .    
\end{array}
\eqno(E.3h)
$$
We shall need the forms of the unperturbed propagators for the unphysical
mesons:
$$
\begin{array}{rcl}
    D^{(0)}_{WW} &=& D_{AA}^{(0)} = D_{ZZ}^{(0)} 
                     = \displaystyle \frac{1}{k^2} \, , \\[1ex]
    D^{(0)}_{AZ} &=& D^{(0)}_{A \phi} =0 \, ,  \\[1ex]
    D^{(0)}_{W \phi} &=& -M^2_0/(k^2)^2 \, ,  \\[1ex]
    D^{(0)}_{Z \phi} &=& -M^{\prime \, 2}_0/(k^2)^2  \, , \\[1ex]
    D^{(0)}_{\phi^+ \phi^-} &=& 
             (k^2-\alpha_W  M^2_0) / (k^2)^2 \, , \\[1ex]
    D^{(0)}_{\phi^0 \phi^0} &=& 
            (k^2-\alpha_Z M^{\prime \, 2}_0) / (k^2)^2 \, .
  \end{array}
\eqno(E.4)
$$
Let us first deal with the $2 \times 2$ mixing matrix given
by (2.2) for the longitudinal~$W$ and the unphysical
charged~$\phi$\@.  Referring to (E.4), we find that the
unperturbed form for this mixing matrix is
\begin{displaymath}
  \left[
  \begin{array}{cc}
    - i \alpha_W / k ^2  
    &  - i\alpha_W KM_0 / (k^2)^2 \\[1.5ex]
    - i\alpha_W KM_0 / (k^2)^2  
    & i (k^2-\alpha_W M^2_0) / (k^2)^2
  \end{array}
  \right] \, .
\end{displaymath}

The inverse of the matrix above is
$$
  \left[
\begin{array}{cc}
      i (k^2-\alpha_W M^2_0) / \alpha_W  
      & i KM_0 \\[1.5ex]
      i K M_0  & -ik^2
\end{array}
  \right]     \, .
\eqno(E.5)
$$
Consequently, the $2 \times 2$ mixing matrix is equal to the inverse of
$$
  \left[
    \begin{array}{cc}
      i (k^2-\alpha_W M^2_0 - \alpha_W \Pi_{WW}) / \alpha_W   
      & i K (M^2_0 + \Pi_{W \phi}) / M_0 \\[1.5ex]
      i K(M^2_0 + \Pi_{W \phi})  / M_0  
      & -i (k^2-\Pi_{\phi^{+} \phi^{-}})
    \end{array}
  \right]      \, .
\eqno(E.6)
$$
Let us impose (E.2a) on the first diagonal matrix
element of the inverse of (E.6).  We get, as in the
alpha gauge of (1.1), the condition (2.5).

With (2.5), we reduce the denominator of the matrix
in (E.5) into
$$
  (k^2)^2 (1-\Pi_{\phi^{+} \phi^{-}}/k^2)/\alpha_W \, ,
\eqno(E.7)
$$
Thus we have
$$
\displaystyle
    D_{\phi^{+} \phi^{-}}(k^2) =
    \frac{k^2-\alpha_W M^2_0 - \alpha_W \Pi_{WW}}
         {(k^2)^2 (1- \Pi_{\phi^{+} \phi^{-}}/k^2)} \, ,
\eqno(E.8a)
$$ 
and
$$
\displaystyle 
    D_{W \phi}(k^2) =
    \frac{M^2_0 + \Pi_{W \phi}}
         {(k^2)^2(1- \Pi_{\phi^{+} \phi^{-}}/k^2)} \, .
\eqno(E.8b)
$$ 
The Ward-Takahashi identity (E.2b) enables us to express
the ghost propagator $D_{\eta^+ \xi^-}$~as
$$
\displaystyle
  D_{\eta^+ \xi^-} (k^2)= \frac{1}{k^2(1+g_0E)} \, .
\eqno(E.9)
$$
The Ward-Takahashi identity (E.2c) relates the 1PI
amplitudes with the vacuum expectation value of the Higgs field.

Next we turn to the $3 \times 3$ mixing matrix for the
longitudinal~$A$, the longitudinal~$Z$, and the unphysical
neutral Higgs meson $\phi^0$\@.  Referring to (E.4), we
find that the unperturbed form of this mixing matrix is
\begin{displaymath}
  \left[
    \begin{array}{ccc}
      - i\alpha_A / k^2  & 0  & 0  \\[1.5ex]
      0  & - i\alpha_Z / k^2  
         & - \alpha_Z KM'_0 / (k^2)^2 \, , \\[1.5ex]
      0  & \alpha_ZKM'_0 / (k^2)^2  
         & i (k^2-\alpha_Z M^{\prime \, 2}_0) / (k^2)^2
    \end{array}
  \right]    \, .
\end{displaymath}
The inverse of the matrix above is 
\begin{displaymath}
  \left[
    \begin{array}{ccc}
      i k^2/ {\alpha_A}  & 0  & 0  \\[1.5ex]
      0  & i (k^2-\alpha_Z M^{\prime \, 2}_0) / {\alpha_Z}  
         & KM'_0  \\[1.5ex]
      0  & -KM'_0   & -ik^2  
    \end{array}
  \right]    \, .
\end{displaymath}
Thus the $3 \times 3$ mixing matrix is equal to the inverse of 
$$
  \left[
    \begin{array}{ccc}
      i \left( k^2 - \alpha_A \Pi_{AA} \right) / \alpha_A  
      & -i \Pi_{AZ}  
      & K \Pi_{A\phi} / M'_0  \\[1.5ex]
      -i \Pi_{AZ}  
      & i \left( k^2 - \alpha_Z M^{\prime \, 2}_0
         - \alpha_Z \Pi_{ZZ} \right) / \alpha_Z  
      & K (\Pi_{Z\phi} + M^{\prime \, 2}_0) / M'_0  \\[1.5ex]
      - K \Pi_{A\phi} / M'_0  
      & - K (\Pi_{Z\phi} + M^{\prime \, 2}_0) / M'_0  
      & -i (k^2 - \Pi_{\phi^0 \phi^0})  
    \end{array}
  \right]    \, .
\eqno(E.10)
$$ 
Imposing (E.3b), we get
$$
\displaystyle
    \Pi_{AZ} \left( 1 - \frac{\Pi_{\phi^0 \phi^0}}{k^2} \right)
    = \Pi_{A \phi} \left(1 + \frac{\Pi_{Z \phi}}
                                   {M^{\prime \, 2}_0} \right) \, .
\eqno(E.11a)
$$ 
By requiring $D_{AA} = 1 / k^2$ and making use
of (E.11a), we get
$$
    \frac{\Pi_{AA}}{M^{\prime \, 2}_0}
    \left(1 -  \frac{\Pi_{\phi^0 \phi^0}}{k^2} \right)
    = \left( \frac{\Pi_{A \phi}}{M^{\prime \, 2}_0} \right)^2 \, ,
\eqno(E.11b)
$$ 
By requiring $D_{AA} = D_{ZZ}$, we get
$$
\displaystyle 
    \left( 1 + \frac{\Pi_{ZZ}}{M^{\prime \, 2}_0} \right)
    \left( 1 - \frac{\Pi_{\phi^0 \phi^0}}{k^2} \right)
    = \left( 1 + \frac{\Pi_{Z \phi}}{M^{\prime \, 2}_0} \right)^2 \, .
\eqno(E.11c)
$$ 
The equations in (E.11) are the same as those
in (3.3).  Thus the relations among the 1PI amplitudes
in the pure alpha gauge defined by the gauge fixing terms
of (1.3) are the same as those in the alpha gauge
defined by the gauge fixing terms of (1.2)
and (1.3).

With (E.11), we find that the determinant of the matrix
in (E.10) is equal to
$$
\displaystyle
    \frac{i(k^2)^3}{\alpha_A \alpha_Z} 
    \left( 1 - \frac{\Pi_{\phi^0\phi^0}}{k^2} \right) \, . 
\eqno(E.12)
$$
The expressions for the propagators are listed in (4.2)
and (4.3).

\end{document}